\documentstyle[12pt,epsfig]{article}
\begin{document}

\begin{titlepage}
\rightline{December 2007}

\vskip 2.5cm
\centerline{\Large \bf Renormalization-scale independence of the physical}
\vskip 0.2cm
\centerline{\Large \bf cosmological constant.}
\vskip 2cm
\centerline{R. Foot$^{a}$, A. Kobakhidze$^{a}$, K. L. McDonald$^{b}$ and
R. R. Volkas$^{a}$\footnote{Electronic address:
rfoot@unimelb.edu.au, archilk@unimelb.edu.au, klmcd@triumf.ca,
raymondv@unimelb.edu.au}}
\vskip 1cm
\centerline{$^{a}$ \ School of Physics, Research Centre for High Energy
Physics,}
\vskip 0.2cm
\centerline{\ The University of Melbourne, Victoria 3010,
Australia}
\vskip 0.4cm
\centerline{$^{b}$ \ Theory Group, TRIUMF, 4004 Wesbrook Mall,
Vancouver, BC V6T 2A3, Canada}

\vskip 4cm
\noindent
Treating the metric as a classical background field, 
we show that the cosmological constant does not run with the renormalization scale -- contrary to
some claims in the literature.

\end{titlepage}

\newpage
It has been argued in the literature that the cosmological constant 
runs with renormalization scale. For a recent review, see Ref.\cite{ccreview} and references therein. 
For example, Ref.\cite{sola}  
takes the standard model potential
\begin{eqnarray}
V = - m^2 \phi^{\dagger} \phi + \frac{\lambda}{2} (\phi^{\dagger} \phi)^2 
\end{eqnarray}
and notes that the classical minimum is
$V_{min} = -\frac{m^4}{2\lambda}$. 
They then argue that since $m$ and $\lambda$ are renormalized parameters defined at a renormalization scale $\mu$, 
that this contribution to the cosmological constant, which they call $\Lambda_{ind}$, runs via
\begin{eqnarray}
\frac{dV_{min}}{dt} = - \frac{d}{dt} \left( \frac{m^4}{2\lambda}\right),~
\end{eqnarray}
where $t=\ln\mu$. 
They also include another contribution, called $\Lambda_{vac}$, which is defined as the part of the cosmological
constant arising from vacuum loops, and evaluate its $\mu$ dependence. Finally they sum the two contributions to obtain 
a $\mu$ dependent cosmological constant.

Actually this proceedure does not include all of the 1-loop corrections
to the cosmological constant since there is an explicit $\mu$ dependent
part of the full 1-loop effective potential which has been ignored. 
However, 
all of the contributions to the cosmological constant 
can be obtained from the minimum
of the full effective potential at any given order in perturbation theory. 
One would expect the effect of the running of the renormalized parameters
in the tree-level potential to be exactly cancelled by the explicit $\mu$ dependence of the 
higher loop contributions.
This is because the effective potential, if properly
defined, is formally renormalization-scale independent. One can easily show that the minimum of the 
potential is also renormalization-scale independent and consequently the total vacuum energy does not 
run with scale. 

In the non-Coleman-Weinberg case (i.e.\ with
$m^2 \neq 0$),
$V_{min}$ is divergent which requires that a bare vacuum energy parameter ($\Lambda_{vac0}$, which we
shall rename $h_0m_0^4$ below) be introduced into the effective
potential. Upon renormalization of the potential, the vacuum energy parameter will run with the 
renormalization scale,
although the vacuum energy, that is the contribution to the cosmological constant, will not.
The essential point is discussed in Ref.\cite{bando,jones}, although these papers did not consider the problem
from the cosmological constant point of view.

Following Ref.\cite{bando}, consider the simple example of $\lambda \phi^4$ model with Lagrangian
\begin{eqnarray}
{\cal L} = \frac{1}{2}  \partial_\mu \phi \partial^\mu \phi - \frac{1}{2} m_0^2 \phi^2 - \frac{1}{4!} \lambda_0 \phi^4 - h_0m_0^4.
\label{bare}
\end{eqnarray}
If one is not interested in the cosmological 
constant contribution, one can just subtract off the divergent vacuum energy 
contributions following the procedure of Ref.\cite{jones}. But for our present purpose,
we have to properly account for the renormalization of the $hm^4$ term.

The effective potential at the 1-loop level expressed in terms of renormalized parameters defined 
in the $\overline{MS}$ scheme is \cite{bando}:
\begin{eqnarray}
V &=& V^{(0)} + V^{(1)} \nonumber \\
V^{(0)} &=& \frac{1}{2} m^2 \phi^2 + \frac{1}{4!} \lambda \phi^4 + hm^4
\nonumber \\
V^{(1)} &=& \frac{1}{64\pi^2} M_{\phi}^4 \left( \ln \frac{M_\phi^2}{\mu^2}  - \frac{3}{2} \right)
\end{eqnarray}
where $M_{\phi}^2 \equiv \frac{1}{2} \lambda \phi^2 + m^2$ and $\mu$ is the renormalization scale.
The relation between the bare ($h_0$) and renormalized ($h$) coupling is
\begin{eqnarray}
h_0 = Z_h \mu^{n-4} h,
\end{eqnarray}
where $n=4-\epsilon$ is the space-time dimension in dimensional regularization. 
Note that the renormalized parameter $h$ becomes dependent on $\mu$, although the bare quantity $h_0$ is,
of course, independent of $\mu$.
Because the effective potential is related to the effective action (it is the zero-momentum term in the momentum expansion)
it must be independent of the renormalization scale $\mu$. This implies the RGE
\begin{eqnarray}
D V (\phi, m^2, \lambda, h, \mu) = 0,
\label{rge3}
\end{eqnarray}
where
\begin{eqnarray}
D = \mu \frac{\partial}{\partial \mu} + \beta \frac{\partial}{\partial \lambda} - \gamma_m m^2 \frac{\partial}{\partial m^2}
- \gamma_\phi \phi \frac{\partial}{\partial \phi} + \beta_h \frac{\partial}{\partial h}\ .
\end{eqnarray}
Of course this RGE is a statement that the $\mu$ dependence of the effective potential 
$V$ due to the 
running of the couplings is exactly cancelled by the
explicit $\mu$ dependence in the potential.

The cosmological constant contribution is set by the value of the effective potential at its minimum.
At the minimum, $\partial V/\partial \phi = 0$ and so Eq.(\ref{rge3}) reduces to:
\begin{eqnarray}
\frac{dV_{min}}{dt} = \mu \frac{\partial V_{min}}{\partial \mu} + 
\beta \frac{\partial V_{min}}{ \partial \lambda} - \gamma_m m^2 \frac{\partial V_{min}}{\partial m^2} 
+ \beta_h \frac{\partial V_{min}}{\partial h}
= 0.
\end{eqnarray}
We see that the vacuum energy is formally scale-invariant.

In the case of the standard model, the effective running of the parameters in the Higgs potential
will be cancelled by the explicit $\mu$ dependence of the full effective potential -- just as in the 
above $\lambda \phi^4$ model. Thus, there is no $\mu$ 
dependence to the cosmological constant contribution. The parameter $h$ does run
of course, but it runs in such a way that the vacuum energy contribution to the cosmological constant is renormalization
scale independent. The running of the tree-level term, $hm^4$, has no physical significance. The physically relevant
quantity is the cosmological constant which, of course, includes all radiative corrections (not just the tree-level
term, $hm^4$). Thus,
equating this tree-level term (or more generally, the minimum of the tree-level potential with
running parameters) to the measured cosmological constant, at the Hubble scale $\mu = H_0 \sim 10^{-33}$ eV,
as done in e.g. ref.\cite{sola}, is not justified.
From the RGE point of view (with  gravity treated as a classical background field) no 
special role is played by the current Hubble scale $H_0 \sim 10^{-33}$ eV.
In particular,
there is no reason that setting $\mu = H_0$ will minimize the loop contributions to the cosmological
constant, so the value of the parameter $h$ at the Hubble scale 
has no special significance. 

The above analysis, which was performed assuming a flat background metric, can in principle be extended 
to generic non-flat background metrics. Despite ambiguities in quantizing fields on curved backgrounds, one can argue that the 
cosmological constant (the non-derivative term in the effective action) does not run. Indeed, the effective gravitational action can 
be defined by 
integrating out quantum ``matter'' fields, collectively denoted here as $\phi$:
\begin{equation}
{\rm e}^{iS_{\rm eff.}[g_{\mu\nu}]}=\int D[\phi]{\rm e}^{iS[\phi, g_{\mu\nu}]}~,
\label{a1}
\end{equation}  
where $S[\phi, g_{\mu\nu}]$ is the bare action. The first term in the derivative expansion 
of the effective action $S_{\rm eff.}$ represents a cosmological constant, $S_{\rm eff.}=\int d^4x\sqrt{-g}(\Lambda_{\rm eff.}+...)$. 
Then scale independence of the 
effective action, $\frac{dS_{\rm eff.}}{dt}=0$, implies scale independence of the effective cosmological constant as well.  

To summarize, we have examined the question of the renormalization scale evolution of 
the cosmological constant. 
We have shown that the cosmological constant does not formally run with renormalization scale, at least
in the case where the gravitational metric is treated as a classical background field.

 \vskip 1cm
 \noindent
 {\bf Acknowledgements:} 
 We thank J. Sola and I. L. Shapiro for correspondence regarding an initial draft of this paper.
 This work was supported by the Australian Research Council.

\end{document}